\documentstyle[12pt]{article}
\begin{document}
\baselineskip=7mm

\newcommand{\TeV}{\,{\rm TeV}}
\newcommand{\GeV}{\,{\rm GeV}}
\newcommand{\MeV}{\,{\rm MeV}}
\newcommand{\keV}{\,{\rm keV}}
\newcommand{\eV}{\,{\rm eV}}

\newcommand{\be}{\begin{equation}}
\newcommand{\ee}{\end{equation}}
\newcommand{\bea}{\begin{eqnarray}}
\newcommand{\eea}{\end{eqnarray}}
\newcommand{\ba}{\begin{array}}
\newcommand{\ea}{\end{array}}
\newcommand{\bmat}{\left(\ba}
\newcommand{\emat}{\ea\right)}
\newcommand{\refs}[1]{(\ref{#1})}
\newcommand{\ler}{\stackrel{\scriptstyle <}{\scriptstyle\sim}}
\newcommand{\ger}{\stackrel{\scriptstyle >}{\scriptstyle\sim}}
\newcommand{\lag}{\langle}
\newcommand{\rag}{\rangle}

\begin{titlepage}
\title{Structure Formation with Majoron Supermultiplet\\
\vspace{-4cm}
\hfill{\normalsize IC/94/306\\}
\hfill{\normalsize hep-ph/9410396 \\}
\hfill{\normalsize  October 1994\\}
\vspace{4cm}}
\author{E.~J.~Chun\thanks{Email address: chun@ictp.trieste.it}\\[0.2cm]
        International Center for Theoretical Physics\\
        P.O.Box 586, 34100 Trieste,  Italy }
\date{}
\maketitle

\begin{abstract}
\baselineskip=7mm
We show that the late-decaying particle scenario may be realized in the
supersymmetric singlet majoron model with the majoron scale $10-200$ TeV.
The smajoron decaying into two neutrinos is the late-decaying particle with
the mass $0.1-1$ TeV and the life-time $2\times10^3-8\times10^4$ seconds.
The lower limit of the majorino mass is $4-40$ TeV in order to avoid the
overclosure of the universe due to the decay-produced LSP.
The muon neutrino and the tau neutrino can be used to explain
the atmospheric and the solar neutrino deficit.
\end{abstract}

\thispagestyle{empty}
\end{titlepage}
\clearpage
\setcounter{page}{1}
\baselineskip=7mm

The conventional cold dark matter (CDM) scenario for the structure
formation predicts more power at small scales in flat inflationary
universe \cite{cobe}.
Among various possibilities to fix this problem,
the late-decaying particle scenario with $\Omega =1$ CDM is based
on the idea that delaying the time of  matter-radiation equality due
to the relativistic decay-products increase  the size of the scale
starting to grow \cite{bbe}. The question now is how this idea can be
realized in a specific particle physics model.
Until now a handful of models have been
appeared.  The first proposal was to use the 17 keV neutrino with the
life-time around 1 yr \cite{be}.
More recently, the author and collaborators suggested a light axino
which decays into the gravitino and the axion in the low-energy
supersymmetry breaking scheme \cite{ckk,kk}.
Another possibility with a massive tau neutrino in a doublet majoron
model with small majoron scale around 20 GeV is worked out in
ref.~\cite{km}.  In a slightly different context, another
intersting suggestion was made in ref.~\cite{dgt} where a heavy tau
neutrino producing an electron neutrino (plus majoron or familon) in
the era of nucleosynthesis is used to fit the power spectrum.
\bigskip

The purpose of this paper is to propose a late-decaying particle scenario
realized in the singlet majoron model \cite{cmp} combined with
supersymmetry.  Conventional ways for fitting the power spectrum is to
use heavy tau neutrinos.
Most well-known is  the mixed dark matter scenario with a tau neutrino
of mass $\sim$ $30h^2$ eV as a hot dark matter component.
Another way is to introduce a heavier tau neutrino of mass 1-10 MeV
mentioned above \cite{dgt}.
This possibility can be realized in the singlet majoron model with
the majoron scale $V_L \simeq 1$ TeV \cite{mn}.
In these cases however  one cannot reconcile both the solar and the
atmospheric
neutrino problem with the minimal number of neutrino species.\footnote{
For a
reconcilation of all the problem one may introduce almost degenerate
neutrinos of masses around a few eV in the mixed dark matter scenario
\cite{degn}.} Introducing one sterile neutrino may provide a model where
an unstable MeV Majorana tau neutrino can reconcile the CDM scenario with
data on solar and atmospheric neutrinos as presented  recently in
ref.~\cite{jv}.

In this paper we do not explain the structure formation by a heavy
neutrino.  Therefore we can have three species of neutrinos with masses
accounting for the deficit of solar and atmospheric neutrinos:
$m_{\nu_\mu} \simeq 10^{-2}-10^{-3} \eV$ and $m_{\nu_\tau}\simeq 0.1 \eV$
\cite{sm}.  In order to provide a good fit for the structure formation
we invoke the late-decaying particle scenario with $\Omega=1$ CDM.
Our scenario uses a particle inherented in the supersymmetric singlet
majoron model: the smajoron $s$ (the scalar partner of the majoron)
which decays into a pair of tau neutrinos.  Supersymmetric
models are endowed with a natural candidate for CDM, namely, the lightest
supersymmetric particle (LSP) when the R-parity is imposed for the proton
stability \cite{nil}.  In our scheme the LSP is assumed preferably to be
the usual neutralinos.  This fact imposes a constraint on the mass of the
fermion partner of the majoron ($J$), called the majorino ($\psi$), since
the majorino decay  produces at least one LSP.
\bigskip

The late-decaying particle scenario assumes a long-lived massive relic
particle $X$ which dominates once the energy density of the universe and
then decays into relativistic particles. Let the mass of $X$ be $m_X$,
the life-time $\tau_X$ and the ratio of the relic number density to the
entropy density $Y_X$.
These relativistic remnants are red-shifted away to delay the time of
the usual matter-radiation equality.
At the time of the new matter-radiation equality,
the length scale $\lambda_{EQ}$ characterizing the evolution of
fluctuation spectrum is given by \cite{kk}
\bea
 \lambda_{EQ} &\simeq& 30\,(\Omega h^2)^{-1}\theta^{1/2}\,\mbox{Mpc}
         \nonumber\\
 \mbox{with}\quad \theta &=&  1 + {b \over 0.67}
         \left(\tau_X \over \sec\right)^{2/3}
         \left( m_X Y_X \over \MeV \right)^{4/3} \,.
\eea
Here $b$ is the fraction of the relativistic energy density from the
$X$ decay.
The shape of the power spectrum is set by the value of
$\kappa = (\Omega h) \theta^{-1/2}$. The best fit requires
$\kappa = 0.2 - 0.3$ \cite{kt}.
Therefore the late-decaying particle scenario with $\Omega =1$ fixes
the basic relation
\be \label{m}
     \left( \tau_X \over \sec\right) \left( m_X Y_X \over \MeV \right)^2
    \simeq  0.55\,b^{-3/2} \left( h^2/\kappa^{2} -1 \right)^{3/2} \;.
\ee
Since the particle $X$  decays  after the nucleosynthesis, their energy
density should be less than that of one neutrino at the time
of the nucleosynthesis. It gives the restriction
\be \label{n}
 {m_X Y_X \over \MeV} < 0.107 \;.
\ee
Then eq.~\refs{m} puts the lower limit on the life-time;
$\tau_X > 114\,b^{-3/2} $, e.g., for $h=0.5$ and $\kappa=0.3$.
What makes the late-decaying particle scenario different from the others
is the existence of an extra small scale corresponding to the horizon at
the first matter-radiation equality.  It is
\be \lambda_{EQ1} \simeq 80 \left( \keV \over m_X Y_X \right)\,
         {\rm kpc.}
\ee
Further investigation is needed to confirm or exclude the existence of
such a scale.
\bigskip

The supersymmetric singlet majoron model in its simplest form
assumes the presence of one extra singlet which couples to three
right-handed neutrinos \cite{gmpr}.  In this model the extra singlet
as well as the right-handed sneutrinos develop non-vanishing vacuum
expectation values of order of the supersymmetry breaking scale
$m_{3/2} \sim 1$ TeV as it is the only scale appearing in the scalar
potential with soft-terms.
Although this kind of model is appealing in its minimality,
it cannot provide a solution to the dark matter problem since the
non-vanishing vacuum expectation value of the right-handed sneutrino
breaks the R-parity and thus destabilizes the LSP.
Therefore one would like to consider other kind of supersymmetric
singlet majoron models where the lepton number breaking scale
$V_L$ is a free parameter \cite{mz}.
The class of models we are considering have the following
superpotential in addition to that of the minimal supersymmetric
standard model:
\be
  W = h_{ij} L_i H N_j + f_{ija} N_i N_j S_a + W'(S_a)
\ee
with three families $N_i$ of right-handed neutrinos and arbitrary
number of singlets $S_a$ \cite{ckl}.  The whole superpotential is
invariant under lepton number,  spontaneous-breaking of which is
provided by the model-dependent superpotential $W'$.
Apart from being larger than $m_{3/2}$ the lepton number breaking
scale $V_L$ is taken as a free parameter.
It is important to observe that, whenever $V_L>m_{3/2}$, the
right-handed sneutrinos do not require a vacuum expectation value
so that the R-parity is not broken \cite{mr}. Therefore the LSP
can be a candidate for cold dark matter.
The masses of the smajoron and the majorino are expected to be smaller or
equal to the supersymmetry breaking scale $m_{3/2}$  \cite{mz,ckl}.
Our late-decaying particle scenario will fix the masses given the scale
$V_L$.
\bigskip

In the models under consideration, strong bounds on $V_L$ were found due
to the fact that the relic density of smajorons (majorinos) can
provide excessive energy density of the universe either at
the time of nucleosynthesis or at present \cite{mz,ckl}.  Smajorons
(majorinos) usually decouple from the thermal bath when they are
relativistic to result in $Y_{s,\psi} \simeq 10^{-3}$.
The relic number density however can be further suppressed due to the
self-couplings among the majoron supermultiplet, which was the key
observation in ref.~\cite{mr}.
In order to see this, we consider the trilinear
coupling of the majoron supermultiplet $\Phi$ to a heavy field $Z$ in the
Higgs superpotential; ${1\over2} \Phi^2 Z + {1\over2} M_Z Z^2$.
The trilinear coupling is absorbed into the (effective) mass $M_Z$.
The six-dimensional  operators arise  in the D-term of
$(\Phi\bar{\Phi})^2/4M_Z^2$ due to the tree-level exchange of $Z$.
In calculating this one can use the supermultiplet formalism
since it involves no supersymmetry breaking effect. And we neglect
contributions due to supersymmetry breaking.
It leads to the following interaction in components:
\be \label{i}
 {\cal L}_{eff} \sim
   {1\over 4M_Z^2}[s^2(\partial_\mu J)^2 + J^2(\partial_\mu s)^2]
       + {1\over 4M_Z^2} J^2\bar\psi_Mi\gamma_\mu\partial^\mu \psi_M
\ee
where $\psi_M$ denotes the majorana spinor of the majorino.
The effective mass $M_Z$ encodes the model-dependence of the
interaction strength.
These interactions are effective even after the decoupling of the smajoron
and the majorino from the thermal bath.  Non-relativistic smajorons
(majorinos)
can follow its thermal distribution to yield the suppressed relic number
density for the values of $V_L$ lower than $10^6$ GeV.

The point now is that smajorons (majorinos) can decay after the ear of
nucleosynthesis without causing any problem if their relic number
$Y_{s,\psi}$ is smaller than $10^{-7}$.  Then for a suitable range of
values for $V_L$ and  $m_s$ ($m_{\psi}$) the condition \refs{m} can be
fulfilled to realize the late-decaying particle scenario.
The supersymmetrized vertex of majoron-neutrino-neutrino also contains
the coupling of smajoron(majorino)-neutrino-(s)neutrino. This will be
the main decay mode of smajoron (majorino) in our case.
The cosmological role of majorinos is different from that of smajorons
in that majorinos produce LSP's which may overclose the universe if
relic number density of majorinos is too much.  As it turns out, our
scheme doesnot allow the majorino to be the LSP contrary to the case in
ref.~\cite{mr}. As we will see, smajorons producing two neutrinos
can properly delay the time of matter-radiation equality for suitable
choices of $V_L$ and $m_s$.  Given $V_L$ and $m_s$ the lower limit
for the majorino mass  has to be put in order to reduce the relic
density in a sufficient amount.
\bigskip

It is now straightforward to calculate the life-time and the relic number
density of smajorons (majorinos) in terms of $m_s$ ($m_\psi$) and $V_L$.
In certain models the five-dimensional  self-interaction among
the majoron supermultiplet are present to cause the decay of the smajoron
into two majorons \cite{ckl}.  Since it leads too fast decay for our
purpose, we assume the absence of this decay mode.  Then producing a pair
of tau neutrinos is the dominant decay mode of the smajoron with the
life-time,
\be
 \tau_s = 1.65 \times 10^2 \left( V_L \over 10\TeV \right)^2
            \left(1\TeV \over m_s \right)
                \left( 0.1 \eV \over m_\nu \right)^2 \sec\;.
\ee
{}From the life-time limit following eq.~\refs{n}, the preferable value of
the scale $V_L$ is bigger than roughly 10 TeV for $m_s \simeq 1$ TeV.

In order to determine the relic number density we compare the interaction
rate with the expansion rate $H$ of the universe.
{}From the above expression \refs{i} one obtains the following interaction
rate \cite{ros};
\bea
  \Gamma_{int} &=& n_s^{\rm EQ} \lag \sigma v_{rel} \rag   \nonumber
        = {m_s^3 \over (2\pi x_s)^{3/2}} e^{-x_s}
                      \sigma_0 (1 + x_s^{-1})  \\
    \mbox{where}\quad \sigma_0 &=& {9 m_s^2 \over 128\pi M_Z^4} \;.
\eea
Here  $x_s \equiv m_s/T_J$ and $T_J$ is the majoron temperature in terms
of which the majoron follows the equilibrium distribution.  The majoron
temperature is related to the photon temperature $T$ :
$T_J = a(T) T$ with $a(T) = [g_{*s}(T)/g_{*s}(T_D)]^{1/3}$ where $T_D$ is
the decoupling temperature of the smajoron out of the (photon) thermal
bath and $g_{*s}$ is the effective degrees of freedom contributing
the entropy density.
Taking  the final decoupling temperature $T_{D}'$ smaller than the
top-quark mass, the reference value of $a(T_D')$ is $a= 0.72$.

The relic population of the smajoron is given by
\be \label{ys}
 Y_s = {45a^3\over (2\pi)^{5/2} \pi g_{*s}(T_D')} x_s^{3/2} e^{-x_s}
\ee
where $x_s$ is determined by
$\Gamma_{int} = H = 1.66 g_*^{1/2} T^2/M_{Pl}$.
The condition for the successful structure formation \refs{m} can now
be analyzed in terms of $m_s$, $\tau_s$ and $Y_s$ given above.

Let us now turn to the question of the majorino mass.
As mentioned earlier the usual neutralinos form  $\Omega=1$
CDM into which majorinos can decay.  Then the relic density  of the
majorino should be suppressed in order for the decay-produced
neutralinos not to overclose the universe.
The number density  $Y_\psi$ is also calculated by equating the
interaction rate for the majorino to the expansion rate of the
universe.  The interaction in eq.~\refs{i} gives
\bea
 \Gamma_{int} &=& {2m_\psi^3 \over (2\pi x_\psi)^{3/2}}
                 e^{-x_\psi} \sigma_0 x_\psi^{-1}  \nonumber\\
 \mbox{where}\quad \sigma_0 &=& {3 m_\psi^2 \over 32\pi M_Z^4} \;.
\eea
The relic number density $Y_\psi$ is two times  $Y_s$ with $x_s$ replaced
by $x_\psi$. For the computations our reference values
are $h=0.5$ and $\kappa=0.3$.  The whole analysis is not sensitive to the
allowed variation of these values.
The condition for the secondary neutralinos not to overclose the universe
reads
\be
  Y_\psi \leq 1.36 \times10^{-11} (h/0.5)^2 \left(60\GeV\over
            m_{\chi^0}\right)
\ee
where $m_{\chi^0}$ is the mass of the LSP.
As one can see below the majorino should be heavier than a few TeV in
order to meet the above restriction.
\bigskip

The analysis shows that our scenario prefers relatively large values
for the smajoron and the majorino mass.
As varing the smajoron mass $m_s$ from 10 GeV to 1 TeV, we get the
values of the majoron scale and the majorino mass as in Table 1.
The presented majorino masses are for the decay-produced LSP
(with $m_{\chi^0}=60$ GeV) to form $\Omega = 1$ CDM.  Varing the
mass of the LSP from 20 to 100 GeV,  these  values increase by a factor
of 0.8 or 1.2, respectively.

As can be seen, the majorino should be heavier than the smajoron by
one to three orders of magnitude.  The discrepancy becomes larger for
smaller smajoron masses (or for smaller $V_L$),
which requires unpleasant tunning of the masses.
We did not show the values for the smajoron mass larger than 1 TeV since
they need the majorino mass too far from the supersymmetry
breaking scale $m_{3/2} \sim 1$ TeV.
Admitting tunning of two orders between the smajoron and the majorino mass
the preferable values are $m_s \simeq 0.1-1$ TeV and $m_\psi \simeq 8-36$
TeV which requires $V_L \simeq 12-36$ TeV.

When the effective interaction becomes stronger ($M_Z < V_L$)
the tunning becomes weaker. For instance, when the mass $M_Z$ is pushed
down up to $M_Z = V_L/10$ we get the results in Table 2.
Again for $m_s \simeq 0.1-1$ TeV we have $V_L \simeq 73-220$ TeV and
$m_\psi \simeq 4-20$ TeV. Therefore an acceptable late-decaying particle
scenario can be realized without too much tunning among the parameters.

For the cases with $m_s = 0.1-1$ TeV,  the smajoron life-time falls
in the range $\tau_s = 2\times10^3-8\times10^4$ sec.
Therefore we have the extra small scale
$\lambda_{EQ1} \simeq 3-21$ kpc.
\bigskip

In conclusion, the late-decaying particle scenario with the smajoron in
the supersymmetric singlet majoron model with $V_L \simeq 10-200$ TeV
is suggested.
One also obtains the lower bound for the majorino mass
since its decay-products contains at least one LSP which may cause the
overclosure of the universe.
The required values for the smajoron and the majorino mass are
$m_s\simeq 0.1-1 $ TeV and $m_\psi \simeq 4-40$ TeV.
Contrary to the  other scenarios with a heavy tau neutrino, our suggestion
can explain both the solar and the atmospheric neutrino deficit
by choosing the appropriate neutrino masses.

\begin{table}
\renewcommand{\arraystretch}{1.5}
\caption{The selected values of the majoron scale $V_L$,
the smajoron mass $m_s$ and the
lower limit of the majorino mass $m_\psi$ realizing the late-decaying
scenario.  The taken values for the parameters are $h=0.5$, $\kappa=0.3$
and $M_Z=V_L$.}\medskip

\begin{tabular}{|l||c|c|c|c|c|c|c|c|c|} \hline
$V_L/1\TeV$   & 4.02 & 7    & 8.63 & 10 & 12 & 13.8 & 20 & 30 & 36.4\\
                 \hline
$m_s/100\GeV$ & 0.1  & 0.322 & 0.5  & 0.681& 1 &1.34 & 2.89 &6.7&10 \\
                      \hline
$m_\psi/1\TeV$& 1.93 & 4.04& 5.34& 6.5& 8.31 &10 &16.4 & 28.1 & 36.4\\
                      \hline
\end{tabular}
\end{table}
\begin{table}
\renewcommand{\arraystretch}{1.5}
\caption{Same as Table 1  but $M_Z=V_L/10$.} \medskip

\begin{tabular}{|l||c|c|c|c|c|c|c|c|c|} \hline
$V_L/1\TeV$  & 24.2 & 40 & 52.4 & 65.0 & 73.2 & 100 & 138 & 170 & 223\\
                          \hline
$m_s/100\GeV$ & 0.1  & 0.285 & 0.5 & 0.782& 1 & 1.91& 3.72 & 5.71 & 10\\
                        \hline
$m_\psi/1\TeV$& 0.982& 1.92 & 2.75 & 3.66 & 4.29 & 6.5 &10 & 13.2 &19 \\
                       \hline
\end{tabular}
\end{table}

\bigskip
{\bf Acknowledgement}: It was a pleasure to make a collaboration
\nolinebreak with J. E. Kim and H. B. Kim on the related work.
The author is grateful to Al.~Yu.~Smirnov for comments on the manuscript.
He would like to thank Professor Abdus Salam, the International
Atomic Energy Agency and UNESCO for hospitality at the International
Center for Theoretical Physics, Trieste.

\end{document}